\documentclass[letterpaper]{JHEP}

\usepackage{graphicx}
\usepackage{psfrag}

\providecommand{\MSbar }{\ensuremath{ \overline{\rm MS} }}
\providecommand{\MSbarbold }{\ensuremath{ \overline{\rm \bf MS} }}
\providecommand{\qqbar }{\ensuremath{ q\overline{q} }}

\providecommand{\ibar}{\ensuremath{\overline{\imath}}}

\makeatletter 
\preprint{hep-ph/0105291 v.\ 2\\
DESY 01--034\\
\@date
}
\makeatother

\title{Subtraction method for NLO corrections in Monte-Carlo 
       event generators for $Z$ boson production}
\author{Yujun Chen$^1$ \hspace{10pt} 
        John Collins$^2$\hspace{10pt} 
        Nadiya Tkachuk$^3$  
        \\
        Physics Department,
        Penn State University, \\
        104 Davey Laboratory,
        University Park PA 16802,
        U.S.A. 
\\ \email{$^1$ ychen@phys.psu.edu }
\\ \email{$^2$ collins@phys.psu.edu}
\\ \email{$^3$ tkachuk@phys.psu.edu}
}

\date{15 June 2001}


\abstract{ We use a subtraction method to construct NLO corrections in 
a Monte-Carlo event generator for the case of vector boson production 
in Drell-Yan processes. Our calculations are carried out both for 
the Bengtsson-Sj{\"o}strand-van Zijl (BSZ) algorithm and for a modified 
algorithm 
proposed by Collins. In the case of the modified algorithm, we compute the
relation between the parton distribution functions and
the ones in the \MSbar{} scheme; this relation 
is the same as the corresponding relation for DIS. 
For the BSZ algorithm, we show that there is no simple relation.
}

\begin{document}

\section{Introduction}
\label{sec:intro}

In an earlier paper by Collins \cite{JC}, a subtraction method was introduced
to consistently take into account next-to-leading order
(NLO) terms for deep-inelastic scattering in a Monte-Carlo event
generator. In that paper, the method was applied to the
photon-gluon-fusion process in an event generator 
that uses the algorithm constructed by
Bengtsson and Sj{\"o}strand (BS) \cite{BSZ} for initial-state showering; 
such event generators are PYTHIA \cite{PYTHIA}, LEPTO \cite{LEPTO} or
RAPGAP \cite{RAPGAP}\footnote{
   The same methods can be applied to event generators like HERWIG that
   use a different algorithm, but the details of the calculation will be
   different. 
}.
In this
paper, we apply this method to massive vector boson production ($W$ or $Z$
production) in hadron-hadron collisions.
The method contrasts with previous methods, e.g., \cite{GM}, at 
incorporating NLO corrections by a reweighting of the events generated by 
showering from the LO matrix elements.
The subtraction method is intended to be applicable to all the
perturbative parts of an event generator, to calculate non-leading
corrections to the hard scattering and the showering.  

As described in \cite{JC}, an event generator
using this subtraction method to correct the hard scattering
generates two classes of events. One
class is obtained from the 
LO parton-model process by showering the initial and final state quarks,
exactly as before. The second class of events is generated by starting
with an NLO subprocess and showering the partons, again exactly as 
before, but with one exception, the exception being that the hard cross
section for the subprocess is equipped with a subtraction that correctly
compensates the double counting between the two classes of events.  This
removes the part of the NLO term that is included in the combination of
the LO parton model and showering, and it cancels the singularity in the
NLO contribution to the cross section.

We will use two different algorithms for the parton showers. One algorithm
is due to Bengtsson, Sj{\"o}strand and van Zijl \cite{BSZ} and is used in
PYTHIA and RAPGAP.  As explained in Ref.\ \cite{JC}, the parton density
functions (pdf's) to be used in the event generator are not those of a
standard scheme, but are specific to the showering algorithm.  We will
actually find that there is no simple relation whatever between the parton
densities to be used with this algorithm and the \MSbar{} pdf's, and that
they are therfore also different from the pdf's needed for the
corresponding algorithm \cite{BS} in DIS.  This happens because of the
changes made by the showering algorithm to the parton model kinematics:
the changes are different from those in DIS, and correlate the kinematics
of the partons from each of the incident hadrons.
The second algorithm is the one defined in \cite{JC} with the specific aim
that the relevant parton kinematic variables do not get changed by the
showering.  For this algorithm, we will show that the pdf's for the
hadron-hadron induced process are the same as in DIS, although both differ
from the \MSbar{} definition.

The NLO corrections for vector boson production come from two different
subprocesses: $gq$ and \qqbar{}.  In this paper we derive the results for $gq$
subprocesses; this is an important correction which is not necessarily
suppressed by the factor of $\alpha_s$ if the gluon density is larger than the
quark densities.  The generalization to \qqbar{} subprocess will 
encounter some complications because of the need to treat soft gluon effects.
So we defer this to the future.

In Sec.\ \ref{sec:algorithm}, we describe the treatment of vector boson
production in an event generator.  Then we compute the effect of combining the
LO cross section with the order $\alpha_s$ part of the shower; this will be
needed as the subtraction term in the NLO calculation.  We present the
calculation for $Z$ bosons, and later, in Sec.\ \ref{sec:W}, specify the
changes needed to treat the production of $W$ bosons.

In Sec.\ \ref {sec:NLO correction}, we carry out the subtraction from the
NLO matrix elements to get the resulting NLO hard differential
cross-sections.  The pdf's in these formulae are not in \MSbar{}
scheme, so we show how to relate them to the ordinary \MSbar{} pdf's.  For
the new algorithm, we find that the relationship between the pdf's is the
same as was found with deep-inelastic scattering in \cite{JC}; while for
the BSZ algorithm we show that no simple relation appears to be possible.

Finally, we summarize and discuss our results in Sec.\
\ref{sec:conclusion}.

\section{Monte-Carlo algorithms}
\label{sec:algorithm}


\subsection{QCD improved parton model and vector boson production}

Our calculations are based on the QCD improved parton model, which
shows that a hard scattering process initiated by two hadrons is the
result of an interaction between one parton (quark or gluon) from each of
the incoming hadrons \cite{text}. 

We consider $Z$ boson production in a collision of two hadrons 
$A$ and
$B$ at a center-of-mass energy $\sqrt{s}$.
The leading-order subprocess is $\qqbar \to Z$.  
As in Fig.\ \ref{fig:drell-yan}, the resulting cross section
$d\sigma^{\rm LO}_{AB}/dy_0$ for producing a $Z$ boson of 
rapidity $y_0$
is obtained by weighting the subprocess cross section with the parton
distribution functions $f_i(x_a)$ and 
$f_{\ibar}(x_b)$, and
summing over all quark-antiquark combinations in the beam and target:
\begin{equation}
  \label{LO}
      \frac{d\sigma^{\rm LO}_{AB}}{dy_0}
     = K \sum_i (A_i^2+V_i^2) f_{i/A}(x_a, M_Z^2)f_{\ibar/B}(x_b, M_Z^2).
\end{equation}
The parton momentum fractions are written as
\begin{equation}
\label{eq:xa.xb}
     x_a = \frac{M_Z}{\sqrt{s}} e^{y_0}, ~~~
     x_b = \frac{M_Z}{\sqrt{s}} e^{-y_0}.
\end{equation}
The parton-level cross-section results in the factor 
$K (A_i^2+V_i^2)$, where
\begin{equation}   
     K = \frac{\sqrt{2} \pi G_F \tau}{3},~~~
     \tau = \frac{M_Z^2}{s} = x_a x_b,
\end{equation}
and 
\begin{equation}
    A_i = T_i^3,\  \ 
    V_i = T_i^3-2Q_i \sin^2\theta_w
\end{equation}
come from the axial and vector couplings in the electro-weak
interaction \cite{text}.  Finally, 
$f_{a/A}(\xi,\mu^2)$ is the 
number density of quarks of
flavor $a$ in hadron $A$ at fractional momentum $\xi$ and a
renormalization/factorization scale $\mu$.

In the usual ``matrix element'' approach, the variable $y_0$ is exactly
the rapidity of the $Z$ boson.  This follows from the approximation of
giving the incoming partons zero transverse momentum and virtuality.
However, in a Monte-Carlo event generator, the quark and antiquark are
given their correct kinematics.  In that case, the variable $y_0$ is not
exactly the rapidity of the $Z$ boson; its precise definition is
$\frac{1}{2}\ln(x_a/x_b)$, where $x_a$ and $x_b$ are the fractional
longitudinal momenta of the incoming partons.  This is a variable that is
well-defined in the generation of a particular event, but that is not
necessarily measurable from the final state of the event.  In the context
of an event generator, Eq.\ (\ref{LO}) must therefore be reinterpreted,
not as the lowest-order approximation to the physical cross section for a
particular $Z$-boson rapidity, but as the cross section for events that
inside the program have a certain value for the variable $y_0$.  Of
course, the motivation for using the variable $y_0$ is that it approaches
the true rapidity $y$ in the limit that the parton transverse momenta and
virtualities approach zero.

Since the $Z$'s decay width is small compared to its mass $M_Z$, 
it is sufficient to compute the production cross sections of the 
bosons. The actually measured cross sections of leptons are computed
by multiplying the cross sections by the appropriate branching ratios.

\FIGURE{
\centering
\psfrag{P_A}{$P_A$}
\psfrag{P_B}{$P_B$}
\psfrag{f_q(x_1)}{$f_i(x_a)$}
\psfrag{f_q(x_2)}{$f_{\ibar}(x_b)$}
\includegraphics [scale=0.55]{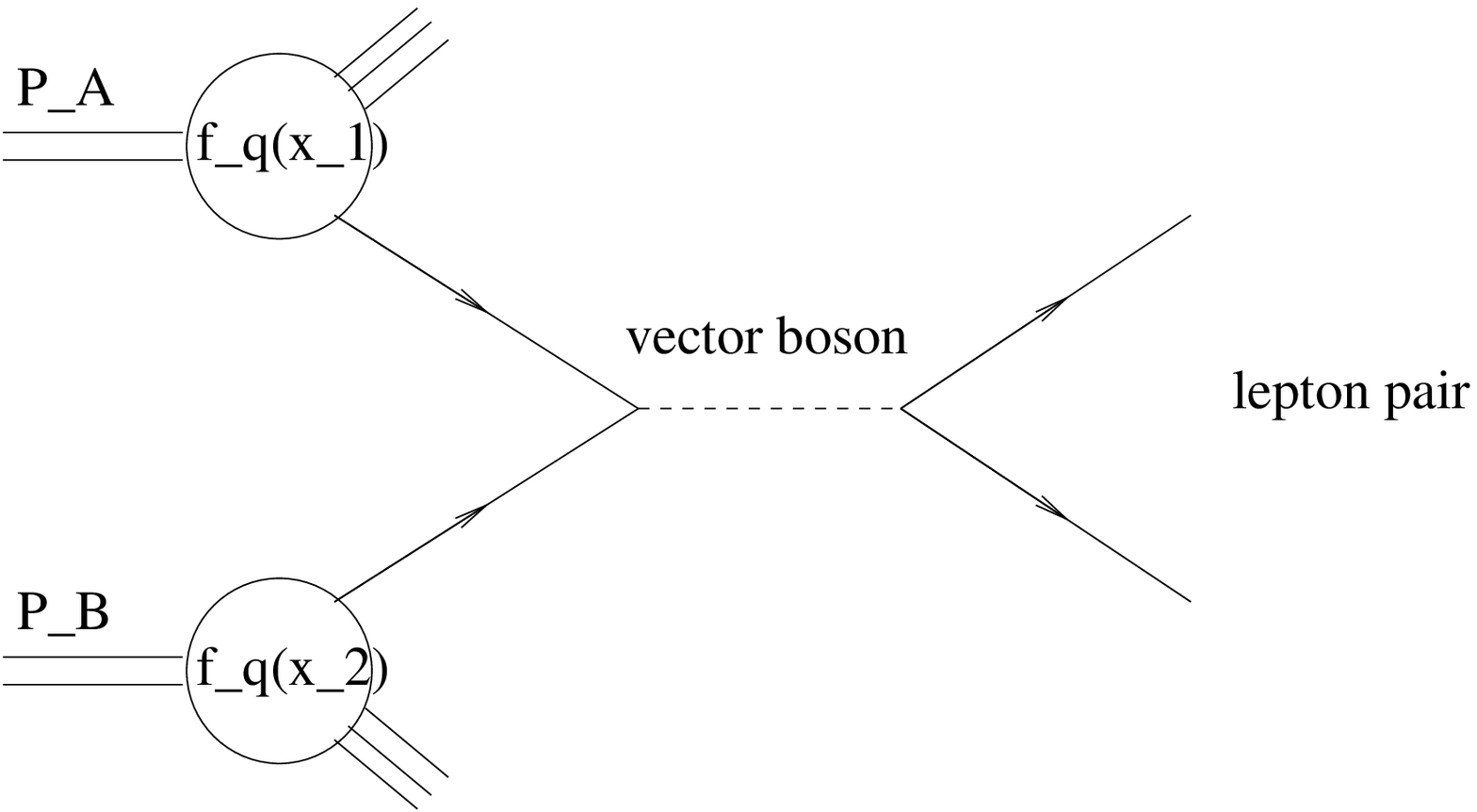}
\caption{Vector boson production in the Drell-Yan model}
\label{fig:drell-yan}
}

With full perturbative QCD corrections taken into account, Eq.\
(\ref{LO}) gets replaced by the factorization formula
\begin{equation}
\label{factorization}
     \frac{ d\sigma_{AB} }{ dy } 
 = \sum_{i,j} \int d\xi_i \, d\xi_j \, 
    f_{i/A}(\xi_i, \mu^2) f_{j/B}(\xi_j, \mu^2) 
        \frac{ d\hat{\sigma}_{ij} }{ dy },
\end{equation}
where $y$ is the rapidity of the $Z$ boson, the $\xi$'s are the momentum
fractions of the incoming partons, and $d\hat\sigma_{ij}/dy$ is a suitably
constructed hard scattering cross section.  Now the sum is over all pairs
of parton flavors (quarks and gluons).  
The formal domain of validity of
Eq.\ (\ref{factorization}) is for the inclusive cross section in the
asymptotic `scaling' limit, analogous to the Bjorken limit in DIS, 
$s\to\infty$ with $\tau$ and $y$ fixed.  As is well-known, 
one effect of
higher-order perturbative corrections is the production of vector bosons
at large transverse momentum ($q_T$).  The discussion above, about 
the distinction
between the variables $y$ and $y_0$, alerts us that care will be needed in
our application of the factorization formula in an event generator.  
In effect, the quantitative interpretation of Eq.\ (\ref{factorization}) gets
modified in current MC event generators; notably the lowest-order formula 
Eq.\ (\ref{LO}) is no longer for $d\sigma/dy$ but
for $d\sigma/dy_0$.  Effectively, when only LO hard scattering 
is concerned and when $q_T$ is small, 
the foundation of an event generator's algorithm is an appropriate 
modification of Eq.\ (\ref{factorization}) that embodies the same physics.
However, when large $q_T$ is concerned, we have to be careful about
the difference between $y$ and $y_0$ for all orders of hard scattering, 
including LO.

\subsection{Parton-shower algorithm}

Now we describe the initial-state shower algorithm for vector boson
production used in PYTHIA, LEPTO or RAPGAP, as described in \cite{BSZ}.

\FIGURE{
\centering
\psfrag{From P_A}{From $P_A$}
\psfrag{From P_B}{From $P_B$}
\psfrag{1}{$1$}
\psfrag{2}{$2$}
\psfrag{1'}{$1'$}
\psfrag{2'}{$2'$}
\psfrag{3}{$3$}
\psfrag{4}{$4$}
\psfrag{3'}{$3'$}
\psfrag{4'}{$4'$}
\psfrag{0 (V)}{0 (Z)}
\includegraphics [scale=0.65]{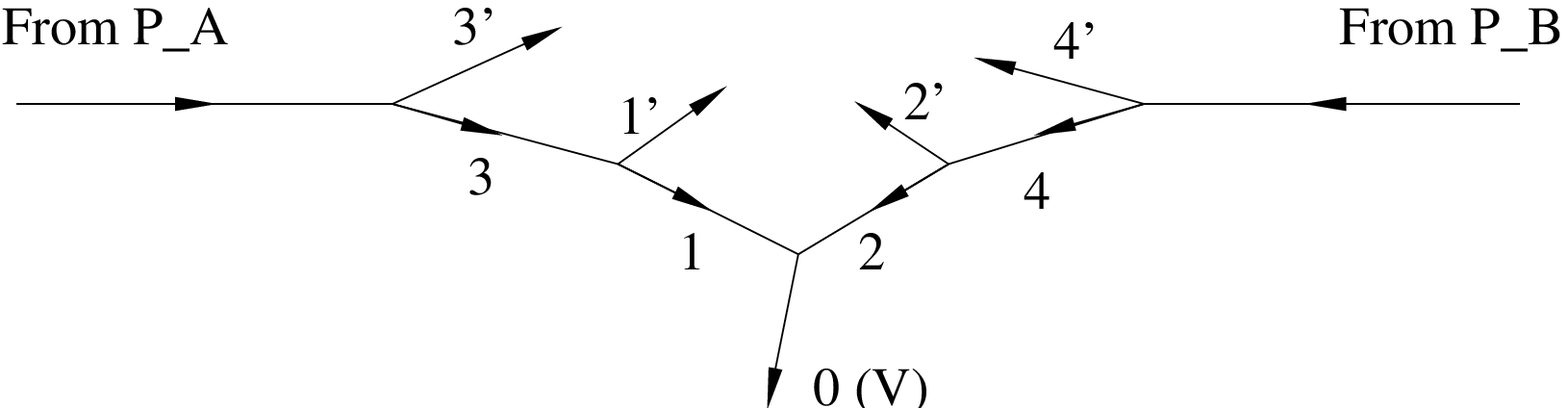}
\caption{Initial-state parton shower for vector boson production}
\label{fig:parton-shower}
}

In Fig.\ \ref{fig:parton-shower} is symbolized an example of
initial-state parton showering for $Z$ production.  The showering algorithm
generates partons with certain flavors and virtualities; it also generates 
the splitting variables $z_{2n+1}$, $z_{2n}$
for each branching, and an azimuthal angle for the transverse momentum of each
branching.  The momentum fraction of each space-like line is computed as
\begin{equation}
\label{x.from.z}
  \xi_{2n+1} = \frac{\xi_1}{\prod_{i=0}^{n-1} z_{2i+1}}, \ \
  \xi_{2n} = \frac{\xi_2}{ \prod_{i=1}^{n-1} z_{2i}},
\end{equation}
so that $\xi_3 = \xi_1/z_1$, $\xi_4 = \xi_2/z_2$, etc.  Since the $z$'s are
generated numerically, by an algorithm explained below, Eq.\ 
(\ref{x.from.z}) gives the values of the momentum fraction variables for
all the lines, except for $\xi_1$ and $\xi_2$.  These values are defined to be
$x_a$ and $x_b$, which are generated according to the probability
distribution corresponding to the lowest order cross section, Eq.\
(\ref{LO}).  

As in \cite{BSZ}, the 
``$\hat{s}$ approach'' is used to relate the splitting variables
$z_i$ to the parton 4-momenta.  This is done by requiring that 
\begin{equation}
\label{s-approach}
   \hat{s}_{ij} = \xi_i \, \xi_j \, s
\end{equation}
both at the hard scattering and at any lower scale in backward showering,
where $\xi_i$ and $\xi_j$ are of the two resolved partons.  This means that
the total $\hat{s}_{ij}$ has to be increased by a factor of $1/z$ in the
backward evolution; this defines the relation between $z$ and parton
kinematics.  For instance, in Fig.\ 
\ref{fig:parton-shower}, if line 1 has the highest virtuality, then
$z_1 = (p_1+p_2)^2/(p_3+p_2)^2$.  

The part of the algorithm used in an event generator that concerns us
is as follows:
\begin{enumerate}

\item Generate values of $y_0$ and $M_Z^2$ from 
the LO cross section for Z boson production Eq.\ (\ref{LO}).
{}From these variables, calculate $x_a$ and $x_b$ by Eq.\ (\ref{eq:xa.xb}).

\item 
   \label{Q1}
   Generate a virtuality $Q_1^2$ for the incoming quark 
   $a$, a
   longitudinal splitting variable fraction $z_1$ for the first branching, and
   an azimuthal angle $\phi$ for this branching. The distributions 
   arise 
   from the Sudakov form factor
\begin{eqnarray}
\label{Sudakov}
   S_i(x_a, Q_{\rm max}^2, Q_1^2) 
   &=& \exp\left\{
      - \int_{Q_1^2}^{Q_{\rm max}^2} \frac{dQ'^2}{Q'^2}
       \frac{\alpha_s(Q'^2)}{2\pi} \right.
\\          
   & & \hspace{27pt} \times
      \sum_k \int_{x_a}^1 \frac{dz_1}{z_1} \, P_{k\to ij}(z_1) 
                        \frac{f_k(x_a/z_1,Q'^2)}{f_i(x_a,Q'^2)}
      \Bigg \}.
\nonumber
\end{eqnarray}
  Here, $Q_{\rm max}^2$ is normally set equal to 
   $M_Z^2$. 
   The Sudakov form factor is the probability that the virtuality of
   quark $a$ is less than $Q_1^2$.

\item Iterate the branching for all initial-state and
   final-state\footnote{
      The final-state showering is organized similarly to the
      initial-state showering.  However, we will not need it
      explicitly in this paper.
   }
   partons until no further branchings are possible.

\item 
   Compute 4-vectors for the momenta of  the generated
   partons.

\end{enumerate}


\subsection{First initial-state branching with BSZ Algorithm}
\label{sec:new.algorithm}
\FIGURE{
\centering
\psfrag{P}{$P_A$}
\psfrag{P_B}{$P_B$}
\psfrag{q}{$q$}
\psfrag{p_1}{$p_1$}
\psfrag{p_2}{$p_2$}
\psfrag{p_3}{$p_3$}
\psfrag{p_1'}{$p_1'$}
\psfrag{s}{$\hat{s}$}
\includegraphics [scale=0.35]{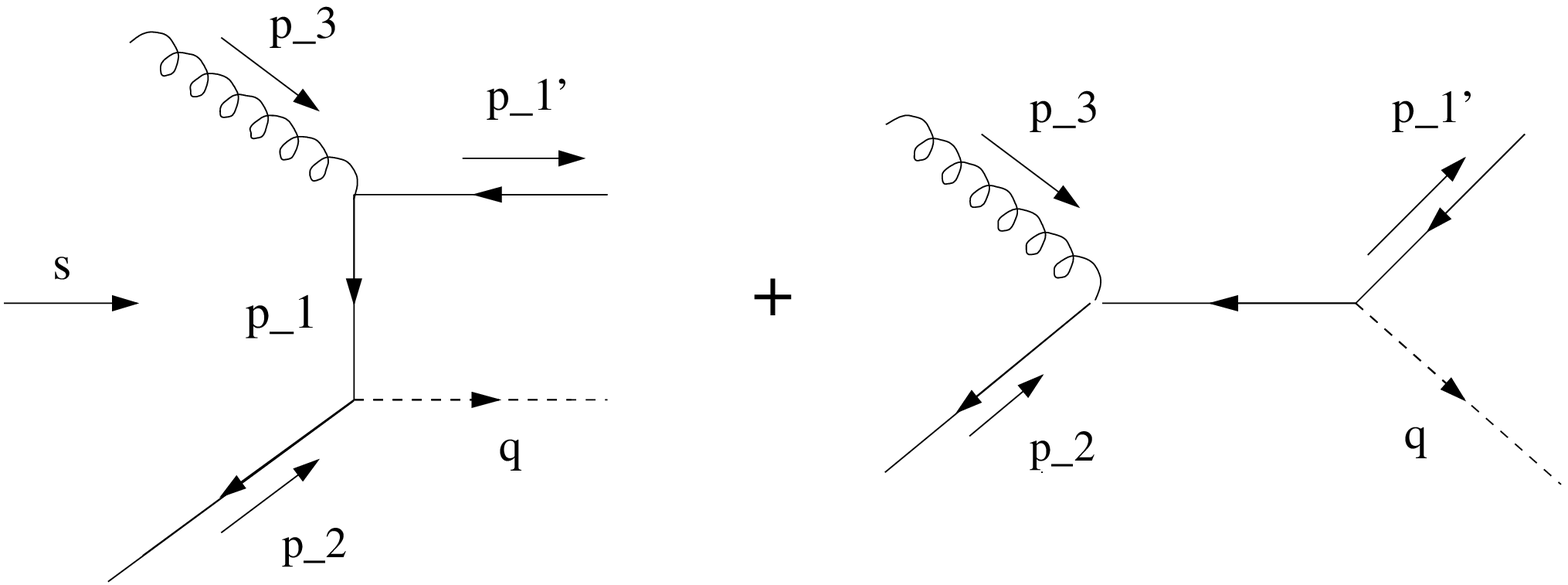}
\caption{NLO gluon-quark collision}
\label{fig:gluon-quark.1}
}

Later, in Sec.\ \ref{sec:NLO correction}, we will calculate the
hard-scattering cross section for the $gq$-induced process, Fig.\ 
\ref{fig:gluon-quark.1}.  The gluon, of momentum $p_3$, comes from
hadron $A$ and the (anti)-quark, of momentum $p_2$, comes from hadron
$B$.  The NLO contribution we want to calculate is to be accurate when
the incoming gluon and quark have virtualities and transverse momenta
that are small compared with $M_Z$, and the intermediate quark,
of momentum $p_1$, has a virtuality of order $M_Z^2$.  To avoid double
counting, it is necessary to subtract the corresponding contribution
obtained from the showering algorithm applied to the LO partonic cross
section,
and it is this subtraction term that we calculate in this section. 

The subtraction is obtained by multiplying the lowest order cross section,
from Eq.\ (\ref{LO}), by the appropriate part of the showering
approximation at order $\alpha_s$.  Only the gluon-to-quark splitting is
relevant for our calculation, and to match with the definition of the NLO
hard-scattering, it should be calculated when the initial-state partons
are on-shell and have zero transverse momentum, and the final-state quark
is on-shell.  The showering factor is just the first order term in the
expansion of the Sudakov form factor~\cite{BSZ} in powers of
$\alpha_s(M_Z^2)$.

The first-order cross section in the showering approximation is
\begin{equation}
\label{Collinear.gq}
   \frac{d\sigma_{\rm shower\, 1}}{dy_0 \, dQ_1^2 \, d \xi_3 \, d\phi}
= K \sum_i ~ (A_i^2 + V_i^2) 
  \frac{\alpha_s(M_Z^2)}{4\pi^2 Q_1^2} \,
       C(Q_1^2) \,
       P(z_1)\,  \frac{ 1 }{ \xi_3 } 
    f_g(\xi_3, M_Z^2) f_{\ibar}(x_b,M_Z^2) .
\end{equation}
Here,  $\xi_3$ is the longitudinal momentum fraction of $p_3$, and 
the splitting kernel is for
$g\to\mbox{quark}+\mbox{antiquark}$: 
$P(z_1) = P_{g\to i\ibar}(z_1) = \frac12 (1-2z_1+2z_1^2)$.  
Because of the way in which an event generator uses the lowest-order cross
section, it is the variable $y_0$ that appears in Eq.\
(\ref{Collinear.gq}) rather than the true rapidity of the $Z$ boson.
We will relate $y_0$ to the true rapidity later. Note 
that because we are doing a strict
expansion in powers of $\alpha_s(M_Z^2)$, the scale argument of the
pdf's is $M_Z^2$.  The function $C(Q_1^2)$ 
is a cut-off function \cite{JC} that gives the maximum value of $Q_1^2$,
and the standard choice is $C(Q_1^2) = \theta(M_Z^2-Q_1^2)$.
We will not discuss other choices of cut-off functions in this paper.

Now we reconstruct the 4-vectors for the momenta $q$, 
$p_1$, $p_3$ and $p_1'$ of the 
vector boson, intermediate quark, the incoming gluon and the outgoing
quark.  In the Bengtsson-Sj{\"o}strand-van Zijl's definition~\cite{BSZ}, 
they obey the following requirements:
\begin{enumerate}
   
\item Hadron $A$ is to be moving in the $z$ direction.

\item Hadron $B$ is to be moving in the $-z$ direction.

\item  The incoming partons, $p_2$ and $p_3$ have momentum fractions
  $\xi_2$ and $\xi_3$ relative to their parent hadrons, in the sense of
  light-front components.  

\item $p_1^2 = -Q_1^2$.

\item 
\label{ext.partons}
  $p_2^2 = p_3^2 = p_1'^2 = 0$, and $p_2$ and $p_3$ have zero
  transverse momentum.

\item 
\label{Z.kinematics.BSZ}
   $q^2 = M_Z^2$, $\xi_2=x_b$.

\item 
\label{z1.BSZ}
   $z_1 = \displaystyle \frac{(p_1+p_2)^2}{(p_3+p_2)^2}=
        \displaystyle \frac{x_ax_bs}{\xi_2\xi_3s} =
        \displaystyle \frac{ x_a }{ \xi_3 } 
      $. 
\end{enumerate}   

In the center-of-mass frame of hadrons $A$ and $B$, with 
the components written in the 
order  $(p^0,{\mathbf p}_{\rm T}, p^z)$, we then have
\begin{eqnarray}
\label{p.1.mu1}
   p_1^\mu &=&
      \left( \frac{1}{2 \sqrt{s}} \left[ \frac{M_Z^2+Q_1^2}{\xi_2}- 
             \frac {Q_1^2}{\xi_3} \right], \mathbf{p}_T, 
             \frac{1}{2 \sqrt{s}} \left[ \frac{M_Z^2+Q_1^2}{\xi_2}+ 
             \frac {Q_1^2}{\xi_3} \right] 
      \right),
\\
\label{p.2.mu1}
   p_2^\mu &=& \frac{\xi_2 \sqrt{s}}{2}(1,\mathbf{0}_T,-1),
\\
\label{p.3.mu}
   p_3^\mu &=& \frac{\xi_3 \sqrt{s}}{2}(1,\mathbf{0}_T,1),
\\
\label{p.1'.mu}
   p_1'^\mu &=& \left( p_3^0 - p_1^0, -\mathbf{p}_T, p_3^z - p_1^z \right),
\\
\label{q.mu1}
   q^\mu &=& \left( p_2^0 + p_1^0, \mathbf{p}_T, p_2^z + p_1^z \right),
\end{eqnarray}
with
\begin{equation}
   \mathbf{p}_T = \sqrt{- \frac{Q_1^4}{\xi_2 \xi_3 s}
                   +Q_1^2 \left( 1- \frac{M_Z^2}{\xi_2 \xi_3 s} \right)}
                   ~\mathbf{n}_T,  
\end{equation}
where $\mathbf{n}_T$ is a unit transverse vector in the direction
defined by the azimuthal angle $\phi$.  

Note that after showering, the rapidity of the $Z$ boson is not
$y_0 \equiv \frac{1}{2}\ln(x_a/x_b)$ but is given by
\begin{equation}
\label{rapidity.exact}
    y\equiv\frac{1}{2} \ln \frac{q^0+q^z}{q^0-q^z} 
     = \frac{1}{2}
        \ln \frac{ \left( Q_1^2+M_Z^2 \right) \xi_3 }
                 { \xi_2 \left( \xi_2\xi_3s-Q_1^2 \right) } ,
\end{equation}
which does approach $y_0$ in the limit $Q_1 \to 0$. 

The natural variables for the LO differential cross section plus the
first-order showering are $Q_1^2$, $y_0$, and $\xi_3$.  However, they are
not so convenient for the NLO corrections.  So we now transform the
cross section in Eq.\ (\ref{Collinear.gq}) in terms of more convenient
variables for a hard gluon-quark scattering: $y$, $\xi_2$, $\xi_3$.

{}From the above equations, we have 
\begin{eqnarray}
\label{eq:Q1}
   Q_1^2 &=& \frac{e^{2y}s \, \xi_2^2  -M_Z^2}{1+e^{2y} \, \xi_2 / \xi_3},
\\
\label{eq:y0}
     y_0 &=& \ln \frac{M_Z}{\sqrt{s}} - \ln \xi_2 ,
\end{eqnarray}
which gives the the Jacobian
\begin{equation}
        \frac{\partial(y_0, Q_1^2)}{\partial(y, \xi_2)}=
        \frac{2e^{2y}\xi_3[\xi_2\xi_3s+M_Z^2]}{(\xi_3+e^{2y}\xi_2)^2} .
\end{equation}
Then the cross section is
\begin{eqnarray}
\label{Collinear.gqBSZ1}
   \frac{d\sigma^{\rm (BSZ)}_{\rm shower\, 1}}{dy \,d\xi_2 \, d \xi_3 \, d\phi}
&=& K \sum_i ~ (A_i^2 + V_i^2) 
  \frac{\alpha_s(M_Z^2)}{4\pi^2 Q_1^2} \,
       C(Q_1^2) \,
       P(z_1)
\\
& & \hspace{35pt}
\times 
    f_g(\xi_3, M_Z^2) f_{\ibar}(\xi_2,M_Z^2)
    \frac{2e^{2y}[\xi_2\xi_3s+M_Z^2]}{(\xi_3+e^{2y}\xi_2)^2} .
\nonumber
\end{eqnarray}


\subsection{First initial-state branching with New algorithm}
\label{sec:BSZ.algorithm}

In the standard algorithms used for treating parton kinematics, the effect
of showering changes the relationship between observable quantities and
the parton momentum fractions from their parton model values.
In Ref.\ 
\cite{JC} an algorithm was proposed that does not suffer from this effect.
Here we extend this algorithm to the Drell-Yan process by requiring 
the rapidity $y$ of the $Z$ boson to be the same after we shower the
incoming partons.  
(Thus it is not necessary to use a separate variable $y_0$ to denote
$\frac{1}{2}\ln(x_a/x_b)$.)

The parton momenta obeys the same requirement as the BSZ algorithm, except
that items \ref{Z.kinematics.BSZ} and \ref{z1.BSZ} are replaced by
\begin{itemize}
\item[$\ref{Z.kinematics.BSZ}'$]
    $q^2=M_Z^2$ and $y=\displaystyle\frac{1}{2}\ln\frac{x_a}{x_b}$.
\item[$\ref{z1.BSZ}'$]
    $z_1 = \displaystyle \frac{ x_a }{ \xi_3 } \not=
        \displaystyle \frac{(p_1+p_2)^2}{(p_3+p_2)^2}$. 
\end{itemize}
Thus the condition on the momentum of $p_2$ is dropped, and instead the
rapidity of the $Z$ boson is required to obey the simple parton-model
relation to $x_a$ and $x_b$.  
In addition, the first splitting variable $z_1$ is defined by \ref{z1.BSZ}$'$,
without using the ``$\hat{s}$'' condition. Accordingly, the fraction momentum
of $p_3$ is required to be $\xi_3=x_a/z_1$ in the new algorithm.  
Note that Eqs.\ (\ref{p.1.mu1})--(\ref{rapidity.exact}) remain true in the
new algorithm.  What has changed is the relation between the parton
kinematics and the variable $x_b$ that is generated by the algorithm.  In
the old algorithm $x_b=\xi_2$; in the new algorithm 
$x_b = x_ae^{-2y} = \xi_3z_1e^{-2y}$.

The first-order cross section in the showering approximation in the new
algorithm is obtained by transforming the cross section in Eq.\ 
(\ref{Collinear.gq}) to be differential in the variables $y$, $\xi_2$,
$\xi_3$, $\phi$, by using Eq.\ (\ref{eq:Q1}), which gives the relation
between $\xi_2$ and $Q_1^2$:
\begin{eqnarray}
\label{Collinear.gq1.New}
     \frac{ d\sigma^{\rm (New)}_{\rm shower \, 1} }
          { dy \, d\xi_2 \, d\xi_3 \, d\phi }
  &=& K  \sum_i ~ (A_i^2+V_i^2) \, \frac{\alpha_s(M_Z^2)}{4\pi^2 Q_1^2}\,
       C_1(Q_1^2)P(z_1)
\\
  && \hspace{35pt}
\times 
       \frac{e^{2y} \left[ \xi_2 \, \xi_3 \, s \, (2+\xi_2/\xi_3 \, e^{2y})
             +M_Z^2 \, \right]}
            {(\xi_3+e^{2y} \, \xi_2)^2} \,
        f_g(\xi_3, M_Z^2) f_{\ibar}(x_b, M_Z^2) ,
\nonumber
\end{eqnarray}
where $y$ is the exact rapidity of $Z$ boson.  Notice that in this
algorithm, although the parton density for the parton of momentum $p_2$ is
calculated with momentum fraction $x_b \equiv e^y M_Z/\sqrt {s}$, the actual
fractional momentum is different (except at zero transverse momentum), and
is given by a rather complicated formula.

This difference between the actual parton momentum fraction and the value
used in the evaluation of the parton density is a characteristic of this
algorithm.  The utility of this apparent inconsistency will become
apparent when we compute the relation with the results of the ``matrix
element'' method of calculation.

\section{NLO hard cross section}
\label{sec:NLO correction}

\subsection{Unsubtracted NLO term}
\label{sec:Unsubstricted}

{}From standard references (e.g., \cite{text}), we find that the
unsubtracted cross section associated with gluon-quark scattering
corresponding to Fig.\ \ref{fig:gluon-quark.1} is:
\begin{eqnarray}
\label{Unsubtracted.gq1}
    \frac{ d\sigma_{\rm unsubtracted \, 1}}
        { d\xi_2 \, d\xi_3 \, d \hat{t} \, d\phi }
&=& \sum_i ~ \frac{\alpha_s(M_Z^2)}{4 \hat{s}^2}
    \frac{\sqrt{2} \, G_F M_Z^2 (A_i^2+V_i^2)}{3(2 \pi)}
\\
&& \hspace{21pt}
\times 
    \frac{\hat{s}^2 + \hat{t}^2 + 2 \hat{u}\, M_Z^2}{  \hat{s} \, \hat{t}}
         f_g(\xi_3,M_Z^2)f_{\ibar}(\xi_2, M_Z^2).
\nonumber
\end{eqnarray}
The sum is over quark and antiquark flavors, and 
\begin{eqnarray}
\label{Mand.variables}
  \hat{s} &=& (p_3+p_2)^2 = \xi_2 \, \xi_3 s,
\\
  \hat{t} &=& -Q_1^2,
\nonumber
\\
  \hat{u} &=& M_Z^2 - \hat{t} -\hat{s} = M_Z^2 + Q_1^2 - \xi_2 \, \xi_3 s.
\nonumber
\end{eqnarray}  
{}From Eq.\ (\ref{eq:Q1}), we have 
\begin{equation}
  \frac {\partial \hat{t}}{\partial y} =
        \frac{-2e^{2y}\xi_2 \, \xi_3 \,[\xi_2 \, \xi_3 \, s + M_Z^2]}
             {(\xi_3 + e^{2y} \xi_2)^2}.
\end{equation}
So now we can also write this unsubtracted cross section in terms of the
same variables as Eq.\ (\ref{Collinear.gq1.New}):
\begin{eqnarray}
\label{Unsubtracted.gq1.a}
     \frac{ d\sigma_{\rm unsubtracted \, 1} }
          { dy \, d\xi_2 \, d\xi_3 \, d\phi }
  &=& K \sum_i ~ (A_i^2+V_i^2) \, \frac{\alpha_s(M_Z^2)}{4\pi^2 Q_1^2}\,
        \frac{\hat{s}^2 + \hat{t}^2 + 2 \hat{u}\, M_Z^2}{\hat{s}^2}
\\
  && \hspace{35pt}
\times 
       \frac{e^{2y}(\xi_2 \xi_3\, s + M_Z^2)}{(\xi_3+e^{2y}\xi_2)^2} \,
        f_g(\xi_3, M_Z^2) f_{\ibar}(\xi_2, M_Z^2) ,
\nonumber
\end{eqnarray}

\subsection{NLO term with subtraction}

We now subtract the showering term, Eq.\ (\ref{Collinear.gqBSZ1}) or
(\ref{Collinear.gq1.New}), from the above $O(\alpha_s)$ term.
A change in
the labeling of the parton  momentum fractions is in order.  The
variables we have previously used, $\xi_2$, $\xi_3$, etc, were tied to a
particular structure for the LO hard scattering and showering.  But
for the subtracted NLO cross section we prefer something that has the
same names for the external partons independently of the subprocess
considered. So we make the following change of notation: Instead of
$\xi_3$, we will write $\xi_a$ to indicate that a parton from hadron $A$,
and instead of $\xi_2$, we will write $\xi_b$ to indicate that a parton
from hadron $B$. 

\begin{itemize}
\item For the BSZ algorithm: 
\begin{eqnarray}
\label{Subtracted.gq1.BSZ}
     \frac{ d\sigma^{\rm (BSZ)}_{\rm hard \, 1}}
          {  dy \, d\xi_a \, d\xi_b \, d\phi }
&=&  K  \sum_i ~ (A_i^2+V_i^2) ~ \frac{\alpha_s(M_Z^2)}{-4\pi^2 \hat{t}} \,
          f_g(\xi_a, M_Z^2) f_{\ibar}(\xi_b, M_Z^2)
\\
&&   \hspace{15pt}
\times
     \frac{e^{2y}[\xi_a \xi_b s + M_Z^2]}{(\xi_a + e^{2y} \xi_b)^2} 
     \Bigg\{
       \frac{ \hat{s}^2 + \hat{t}^2 + 2 \hat{u}\,  M_Z^2 }
            {\hat{s}^2}- 2 C_1(-\hat{t})\, P(z_1) 
      \Bigg \}
\nonumber
\end{eqnarray}
\item For the new algorithm:
\begin{eqnarray}
\label{Subtracted.gq1.New}
   \frac{ d\sigma^{\rm (New)}_{\rm hard \, 1}}
        {  dy \, d\xi_a \, d\xi_b \, d\phi }
&=&     K  \sum_i ~ (A_i^2+V_i^2) ~ \frac{\alpha_s(M_Z^2)}{-4\pi^2 \hat{t}} \,
           \frac{ e^{2y} }{ \left( \xi_a+e^{2y} \xi_b \right)^2 } \,
           f_g(\xi_a, M_Z^2)
\nonumber\\
&&  \hspace{20pt}
\times 
    \left\{
       f_{\ibar}(\xi_b, M_Z^2)\, 
       \frac{ \hat{s}^2 + \hat{t}^2 + 2 \hat{u}\,  M_Z^2 }
            { \hat{s}^2 }\, \left( \hat{s} + M_Z^2 \right)
    \right.
\\
&&     ~ \hspace{30pt}
   -  f_{\ibar}(x_b, M_Z^2) \, C_1(-\hat{t})\, P(z_1) \, \hat{s}\,
       \left( 2 + \frac{\xi_b}{\xi_a} \, e^{2y} + \frac{M_Z^2}{\xi_a \xi_b s} 
       \right)
     \Bigg \}.    
 \nonumber
 \end{eqnarray}
\end{itemize}
In both equations $\hat{s}$, $\hat{t}$, $\hat{u}$, 
and $z_1$ are all functions of $\xi_a$ and 
$\xi_b$, given by the formulae earlier, which in terms of the new
notation are:
\begin{eqnarray}
   - \hat{t} &=& Q_1^2 
    = \frac{e^{2y}s \, \xi_b^2  -M_Z^2}{1+e^{2y} \, \xi_b / \xi_a}\,,
\\   -\hat{u} &=& \xi_a \xi_b s - M_Z^2 - Q_1^2\,,
\\
   z_1 &=& \frac{x_a}{\xi_a}.
\end{eqnarray}

\subsection{Results for $gq$ subprocess}
\label{sec:gq2-subprocess}

\FIGURE{
\parbox{\textwidth}{~~~}
\centering
\psfrag{P}{$P_A$}
\psfrag{P_B}{$P_B$}
\psfrag{q}{$q$}
\psfrag{p_1}{$p_1$}
\psfrag{p_2}{$p_2$}
\psfrag{p_2'}{$p_2'$}
\psfrag{p_4}{$p_4$}
\psfrag{s}{$\hat{s}$}
\includegraphics [scale=0.35]{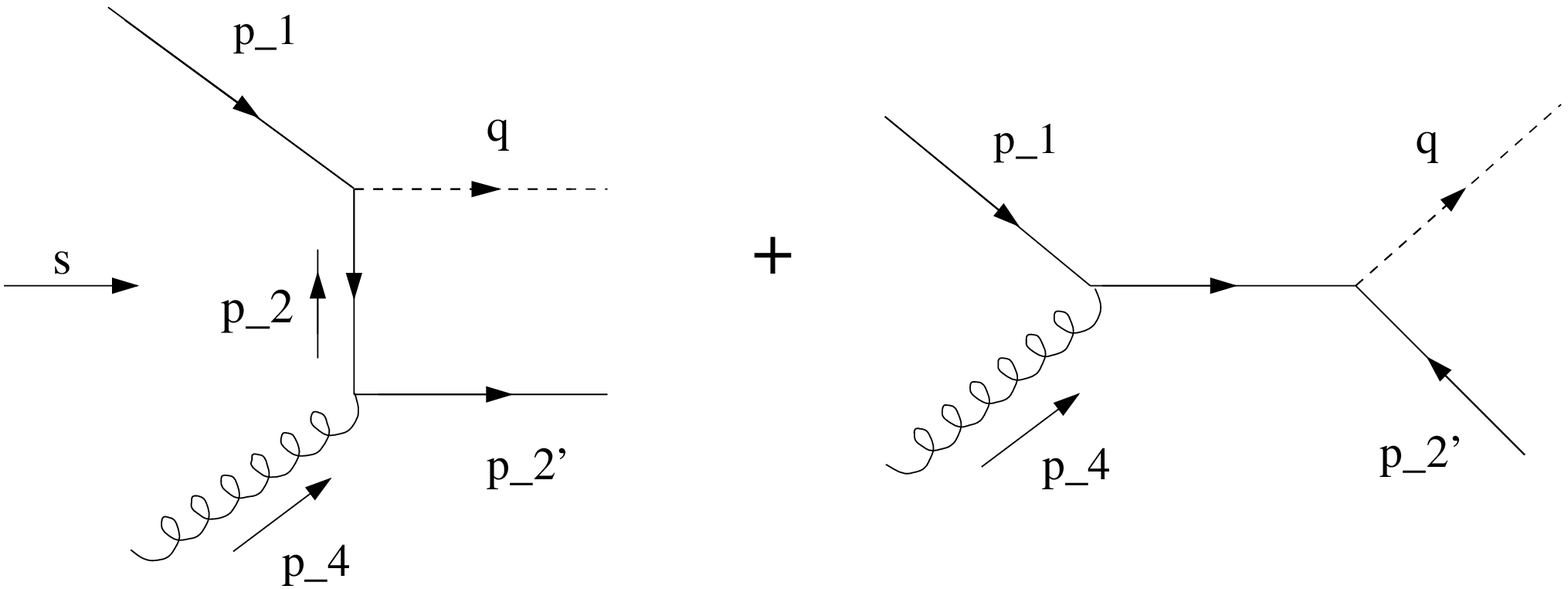}
\caption{NLO gluon-quark collision}
\label{fig:gluon-quark.2}
}

Now we present the results for the other gluon-quark scattering subprocess, in
which gluon comes out of hadron $B$ instead of $A$, from diagrams
shown in Fig.\ \ref{fig:gluon-quark.2}.  The formulae for the
subtracted cross section are obtained from Eqs.\
(\ref{Subtracted.gq1.BSZ}) and (\ref{Subtracted.gq1.New}) by 
making the following change: 
\begin{equation}
  i \leftrightarrow \ibar, ~ a \leftrightarrow b, ~ y \to -y. 
\nonumber
\end{equation}
The resulting subtracted NLO cross sections are:
\begin{itemize}
\item For the BSZ algorithm: 
\begin{eqnarray}
\label{Subtracted.gq2.BSZ}
     \frac{ d\sigma^{\rm (BSZ)}_{\rm hard \, 2}}
          {  dy \, d\xi_a \, d\xi_b \, d\phi }
&=&  K  \sum_i ~ (A_i^2+V_i^2) ~ \frac{\alpha_s(M_Z^2)}{-4\pi^2 \hat{u}} \,
          f_{i}(\xi_a, M_Z^2) f_g(\xi_b, M_Z^2) 
\\
&&   \hspace{15pt}
\times
     \frac{e^{-2y}[\xi_a \xi_b s + M_Z^2]}{(\xi_b + e^{-2y} \xi_a)^2} 
     \Bigg\{
       \frac{ \hat{s}^2 + \hat{u}^2 + 2 \hat{t}\,  M_Z^2 }
            {\hat{s}^2}- 2 C_2(-\hat{u})\, P(z_2) 
      \Bigg \} .
\nonumber
\end{eqnarray}
\item For the new algorithm:
\begin{eqnarray}
\label{Subtracted.gq2.New}
   \frac{ d\sigma^{\rm (New)}_{\rm hard \, 2} }
        {  dy \, d\xi_a \, d\xi_b \, d\phi }
&=& K  \sum_i ~ (A_i^2+V_i^2) ~ \frac{\alpha_s(M_Z^2)}{-4\pi^2 \hat{u}} \,
           \frac{e^{-2y}}{ \left( \xi_b+e^{-2y} \xi_a \right)^2 } \,
           f_g(\xi_b, M_Z^2)
\nonumber\\
&& \hspace{10pt}
\times 
   \left\{
     f_i(\xi_a, M_Z^2) 
     \frac{ \hat{s}^2 + \hat{u}^2 + 2 \hat{t}\,  M_Z^2 }
          { \hat{s}^2 }\, 
     \left( \hat{s} + M_Z^2 \right)
  \right.
\\
&& ~ \hspace{15pt}
   -  f_i(x_a, M_Z^2) \, C_2(-\hat{u})\, P(z_2) \, \hat{s}\,
      \left (2 + \frac{\xi_a}{\xi_b} \, e^{-2y} + \frac{M_Z^2}{\xi_a \xi_b s} 
      \right)
  \Bigg \} .
\nonumber
\end{eqnarray}
\end{itemize}
Here, $z_2 = x_b/\xi_b$.

\subsection{Comparison with \MSbarbold{} scheme}
\label{sec:scheme}

As discussed in \cite{JC}, after we have obtained the NLO corrections,
it is necessary to find the relation between the
scheme of the pdf's used in the event generator and the
commonly used \MSbar{} scheme.\footnote{
   The need for the change of scheme is not apparent in other work on
   merging parton showers and matrix elements \cite{GM}.  Our work relies
   on a deeper analysis of the derivation of the whole algorithm used in a
   Monte-Carlo event generator, rather than just a consideration of the
   normalization of a particular inclusive cross section.  It is only in
   this context that it becomes apparent that the \MSbar{} parton
   densities are not the appropriate ones.  The parton densities used in
   an event generator implicitly include observed jets that result from
   initial-state showering, and therefore the definition of the parton
   density is directly tied to the definition of the showering algorithm. 
   One way to ensure that the pdf's for the event generator are
   the same as \MSbar{} pdf's is to adjust the cut off function $C$
   suitably, as is done by P\"otter \cite{Poetter}.
}
As in \cite{JC} we will compare the
same cross section computed with the \MSbar{} pdf's and
with the formulae used in the Monte-Carlo approach.  We will choose to
calculate $d\sigma/dy$, since 
\begin{itemize}

\item This cross section can be calculated analytically both in the normal
  factorization method and from the algorithm used in the event
  generator.  (The inclusive cross section allows an integral over the
  showering probabilities.)

\item The lowest order has a factorized dependence on both parton momentum
  fractions, so there is exactly enough information to extract the parton
  densities. 

\end{itemize}
As already discussed in \cite{JC}, the unsubtracted part of the cross
section is the same in both methods of calculation, so that the
result of the calculation depends only on the subtraction terms.  The
relation between the pdf's is therefore independent of the
details of the electro-weak couplings for example. 

It is less obvious that the relation between the pdf's could be
unaffected by the differences between the kinematics for the DIS and DY
processes.  This is what we will find, nevertheless.  

In order to get a complete determination of the parton densities, we
consider a modified cross section, in which only a single quark and
antiquark flavor annihilate, and in which $M_Z$ is given an arbitrary
value.  Moreover, the electroweak couplings form a common factor, and it
is convenient to remove them and to arrange that the lowest order cross
section is just $f_i(x_a) f_{\ibar}(x_b)$; this we call the structure
function $\mathcal{F}_i(y, M_Z^2)$.  First we take the formula for
the structure function in the \MSbar{} scheme \cite{SMRS}:
\begin{eqnarray}
\label{F.MSbar}
  \mathcal{F}_i(y, M_Z^2)
&=&     
   f^{(\MSbar)}_i(x_a, \mu^2)f^{(\MSbar)}_{\ibar}(x_b, \mu^2)
\nonumber\\
&& \hspace*{-1.5cm}
 +  \frac{\alpha_s(\mu^2)}{2\pi}
        \int_{x_a}^1 d\xi_a  \int_{x_b}^1 d\xi_b      
   \Bigg \{
        f^{(\MSbar)}_g(\xi_a, \mu^2)
\nonumber\\
&&  \hspace*{-1.2cm} \times
     \Bigg [ f^{(\MSbar)}_{\ibar}(x_b, \mu^2) \,
        \delta(\xi_b-x_b)\, \frac{1}{\xi_a}    
       \Bigg ( P(z_1) 
        \ln \frac{2 (\xi_a-x_a) (1-x_b)}{x_b(\xi_a+x_a)}
        +z_1(1-z_1) \, \Bigg )
\nonumber\\
&& \hspace*{-0.5cm}
    + f^{(\MSbar)}_{\ibar}(\xi_b, \mu^2) 
      \left(
             \frac{G^c(\xi_a,\xi_b)}{\xi_b - x_b} + H^c(\xi_a,\xi_b)\,
      \right)
    - f^{(\MSbar)}_{\ibar}(x_b, \mu^2) 
      \frac{G^c(\xi_a,x_b)}{\xi_b - x_b}
    \Bigg ]
\nonumber\\
&& \hspace*{-1cm}
    + \left( \ibar \leftrightarrow i,\, a \leftrightarrow b, \,
        y \leftrightarrow -y \right)\,
\Bigg \}
   + \mbox{first-order quark terms} + O(\alpha_s^2)  
\end{eqnarray}
with 
\begin{equation}
\label{G-function}
    G^c(\xi_a, \xi_b) = 
      \frac{x_b (\tau + \xi_a \, \xi_b) 
          \left[ \tau^2 + (\tau - \xi_a \, \xi_b)^2 \right] }
           {\xi_a^3 \, \xi_b^2 \, (\xi_a \, x_b + \xi_b \, x_a)(\xi_b+x_b)},
\end{equation}
and
\begin{equation}
\label{H-function}
  H^c(\xi_a, \xi_b) = 
      \frac{\tau (\tau + \xi_a \, \xi_b) 
            \left[ \xi_a \, \xi_b^2 \, x_a
                  + \tau (\xi_a x_b + 2 \, \xi_b \, x_a) \right] }
           { (\xi_a \, \xi_b)^2 \, (\xi_a \, x_b + \xi_b \, x_a)^3}.
\end{equation}
Here the variables $x_a$ and $x_b$ are defined to be exactly the
``parton-model'' values $x_a=\sqrt{\tau}e^y$ and $x_b=\sqrt{\tau}e^{-y}$,
with $\tau=M_Z^2/s$.  The variable $z_1$ in the splitting kernel is
defined as $z_1= x_a/\xi_a$, the same as in the new algorithm.

This structure function must equal the same structure function given by the 
Monte-Carlo calculation.  Since the expressions from the new algorithm are
simpler and have similar structure, we will first compare with the
structure obtained using the new algorithm .

We start from Eq.\ (\ref{Subtracted.gq1.New}) for the cross section.
Using the functions defined in Eq.\ (\ref{G-function}) and Eq.\ 
(\ref{H-function}), we rewrite so as to obtain a formula with a
similar structure to the \MSbar{} formula:
\begin{eqnarray}
\label{F.New}
   \mathcal{F}_i(y, M_Z^2)
&=& 
   f_i^{\rm (new)}(x_a, M_Z^2)f_{\ibar}^{\rm (new)}(x_b, M_Z^2)
\nonumber\\
&&  
   +  \frac{\alpha_s(M_Z^2)}{2\pi}
      \int_{x_a}^1 d\xi_a  \int_{x_b}^1 d\xi_b       
      \Bigg \{ 
        f_g^{\rm (new)}(\xi_a, M_Z^2)
        f_{\ibar}^{\rm (new)}(x_b, M_Z^2)
\nonumber\\
&& 
      \times
        \delta(\xi_b-x_b)
        C_1(Q_1^2)\, P(z_1) 
        \frac{\xi_b^2 \, (\xi_a^2 - x_a^2) - (\tau + \xi_a \, \xi_b)^2}
             {x_a \xi_a\, (\xi_b^2-x_b^2)\,(\xi_a x_b + \xi_b  x_a)}
\nonumber\\ 
&& 
    + f_g^{\rm (new)}(\xi_a, M_Z^2)
      f_{\ibar}^{\rm (new)}(\xi_b, M_Z^2)
     \Bigg [ \, 
      \frac{G^c(\xi_a,\xi_b)}{\xi_b - x_b} + H^c(\xi_a,\xi_b)\,
     \Bigg ]
\nonumber\\ 
&& 
     +  ( \ibar \leftrightarrow i,\, a \leftrightarrow b, \, y 
        \leftrightarrow -y )\,
     \Bigg \} 
\nonumber\\
&& 
     +  \mbox{first-order quark terms} + O(\alpha_s^2)
\end{eqnarray}

Comparison of Eq.\ (\ref{F.New}) and Eq.\ (\ref{F.MSbar}) shows that 
\begin{eqnarray}
\label{pdf.relation.new}
  x f^{\rm (new)}_a(x, M_Z^2)
  &=&  x f^{(\MSbar)}_a(x, M_Z^2)
\nonumber\\
   && + \frac{ \alpha_s(M_Z^2) }{ 2\pi }
       \int_x^1 d\xi  
       \frac{x}{\xi} f^{(\MSbar)}_g(\xi, M_Z^2) 
        \left[
          P(z) \ln (1-z) + z (1-z)
       \right]
\nonumber\\
   && 
   + \mbox{first-order quark terms} + O(\alpha_s^2),
\end{eqnarray}
and $z = x/\xi$,
the same as in Ref.\ \cite{JC}.

When we attempt a similar calculation with the BSZ algorithm, we find that
the parton densities in the subtraction term in the equivalent of Eq.\ 
(\ref{F.New}) are no longer a product of $f_g(\xi_a)$ and $f_{\ibar}(x_b)$.
Instead the argument, $x_b=\sqrt{\tau}e^{-y}$, of the second parton density
is replaced by a complicated function of the kinematic variables.  This
does not match the structure of the \MSbar{} formula Eq.\ (\ref{F.MSbar}),
and it is not possible in any simple way to extract a relation between the
parton densities for the BSZ algorithm and for the \MSbar{} scheme.  The
problem is that the showering of the parton on the $A$ side has affected
the kinematics of the parton on the $B$ side and that it is the modified
kinematics that are used in the corresponding parton density.  As far as
we can see, a correct analysis can only be done by investigating the
problem in terms of unintegrated parton densities.  Since this is a much
more complicated problem, we shall not attempt it here.


\section{$W$ production}
\label{sec:W}

In the above sections we described only $Z$ production since the formulae are 
for annihilation of a quark $i$ with its antiquark $\ibar$.  For $W$ 
production the results are similar, only that one needs a 
different flavor of antiquark $\ibar'$ and an appropriate change in the 
overall coupling: 
\begin{equation}       
  \sqrt{2}\, G_F \, (A_i^2+V_i^2)\, M_Z^2 \longrightarrow
  \sqrt{2}\, G_F \, |V_{i i'}|^2 \, M_W^2,
\end{equation}
where $V_{i i'}$ is an element of CKM matrix.


\section{Conclusion}
\label{sec:conclusion}

We showed how to incorporate the gluon-quark processes in a Monte-Carlo
event generator using the subtraction method proposed by Collins in Ref.\
\cite{JC}. 
We also analyzed the exact parton kinematics used in the BSZ algorithm,
and observed that the factorization theorem for the Drell-Yan process is
used in a modified form compared with the form normally used for the
inclusive cross section $d\sigma/dy$.

When we computed the relation between the parton densities for
the event generator and the \MSbar{} densities, we found the same relation
as found in \cite{JC} in the context of deep-inelastic scattering, but
only if we used the new algorithm for parton kinematics that was proposed
in \cite{JC}.  This supports the hypothesis of process independence of the
pdf's. 
For the case of the normal BSZ algorithm, we found that the effect of the
correlated parton kinematics appears not to permit us to obtain a simple
relation.

\acknowledgments

This work was supported in part by the U.S.\ Department of Energy
under grant number DE-FG02-90ER-40577.
JCC is grateful to the Alexander von Humboldt foundation for an award. 
We would like to thank Xiaomin Zu for discussions.



\begin{thebibliography}{99}

\bibitem{JC}
   J. Collins, JHEP 05 (2000) 004,
   \hepph{0001040}.

\bibitem{BSZ} 
   T. Sj{\"o}strand, \plb{157}{1985}{321} ;
\\
   M. Bengtsson, T. Sj{\"o}strand and M. van Zijl, \zpc{32}{1986}{67}.

\bibitem{PYTHIA}
   T.~Sj{\"o}strand,
   \cpc{82}{1994}{74}.

\bibitem{LEPTO} 
   G. Ingelman, A. Edin and J. Rathsman, \cpc{101}{1997}{108}, 
   \hepph{9605286}.

\bibitem{RAPGAP}
   H. Jung, \cpc{86}{1995}{147}.

\bibitem{GM}
   G. Miu and T. Sj{\"o}strand, \plb{449}{1999}{313},
   \hepph{9812455};
\\
   G. Miu, LU--TP--98--9,
   \hepph{9804317}.
\\
   G. Corcella and M.H. Seymour,  \npb{565}{2000}{227},
   \hepph{9908388}.

\bibitem{BS}
   M. Bengtsson and T. Sj{\"o}strand, \zpc{37}{1988}{465}.

\bibitem{text}
   R. Field, {\it Applications of Perturbative QCD}, Addison-Wesley 1989;
\\
   R.K. Ellis, W.J. Stirling and B.R. Webber, {\it QCD and Collider
   Physics}, Cambridge University Press 1996.

\bibitem{Poetter}
   B.~P\"otter,
   \prd{63}{2001}{114017}, 
   \hepph{0007172}.

\bibitem{SMRS} 
   P.J. Sutton, A.D. Martin, R.G. Roberts and W.J. Stirling, 
   \prd{45}{1992}{2349};
\\
   P.J. Rijken, W.L. van Neerven, \prd{51}{1995}{44},
   \hepph{9408366}.

\end{thebibliography}
\end{document}